\documentclass[letterpaper, 10 pt, conference]{ieeeconf}  

\IEEEoverridecommandlockouts                              

\usepackage{cite}

   \usepackage{graphicx}

\usepackage{amsmath}

\usepackage{algorithmic}

  \usepackage[caption=false,font=footnotesize]{subfig}

\usepackage{stfloats}

\usepackage{url}

\usepackage{tikz}
\usepackage{tikzscale}
\usepackage{tubscolors}
\usetikzlibrary{shapes,automata,arrows,matrix,backgrounds,fit,patterns,decorations.markings, svg.path, shapes.multipart, external, shadows, positioning, calc, decorations}
\tikzstyle{shadedBlue} = [top color=tuDarkBlue20, bottom color=tuDarkBlue40, draw=tuDarkBlue80, thick]%
\tikzstyle{shadedRed} = [top color=tuRed20, bottom color=tuRed40, draw=tuRed80, thick]%
\tikzstyle{shadedOrange}  = [top color=white, bottom color=tuOrange40,  draw=tuOrange100, thick]%
\tikzstyle{shadedGreen}  = [top color=white, bottom color=tuGreen40,  draw=tuGreen100, thick]%
\tikzstyle{shadedYellow} = [top color=white, bottom color=tuYellow40,  draw=tuYellow100, thick]%
\tikzstyle{shadedGray} = [top color=white, bottom color=tuGray20, draw=tuGray60, thick]%
\tikzstyle{shadedGrayLight} = [top color=tuBlack!5, bottom color=tuBlack!10, draw=tuBlack!30, thick]%
\tikzstyle{shadow} = [drop shadow={opacity=.5,shadow xshift=.3ex,shadow yshift=-.3ex}]%
\tikzstyle{triangle} = [isosceles triangle,isosceles triangle stretches]%
\tikzstyle{label-it} = [font=\itshape]%
\tikzstyle{block} = [draw, shadedBlue, rectangle, rounded corners, minimum height=2em, minimum width=5em]%
\tikzstyle{smallblock} = [draw, shadedBlue, rectangle, rounded corners, minimum height=1em, minimum width=2em,shadow]%
\tikzstyle{outerblock} = [draw, shadedGrayLight, draw=tuGray80, rectangle, rounded corners, minimum height=2em, minimum width=5em,shadow]%
\tikzstyle{bubble} = [fill=black,shadow,circle,draw=black,inner sep=0pt,minimum size=5pt]%
\tikzstyle{memory} = [cylinder, shape border rotate=90, aspect=.4, shadedGrayLight, minimum width=5em, shadow]%
\tikzstyle{inheritArrow} = [-open triangle 60,thick]%
\tikzstyle{kompArrow}    = [diamond-,thick]%
\tikzstyle{flowDecision} = [diamond, draw, shadedRed, text badly centered, inner sep=0pt,shadow]%
\tikzstyle{flowBlock} = [rectangle, draw, shadedBlue, text centered, rounded corners, minimum height=2em,shadow]%
\tikzstyle{bgBox} = [rectangle, draw, shadedGrayLight, text centered, rounded corners=3mm, shadow, inner sep=10pt]%
\tikzstyle{blockarrow} = [draw, thick, single arrow, minimum height=3em]%
\tikzexternaldisable

\usepackage{pgfplots}
\usepackage{pgfplotstable}

\usepackage{graphicx} 
\usepackage{xr}
\usepackage{amssymb} 	
\usepackage{bm} 		
\usepackage{siunitx}	
\usepackage{booktabs} 	
\usepackage{multirow}   
\usepackage{colortbl}
\usepackage{array}
\usepackage{lmodern}

\newcolumntype{P}[1]{>{\centering\arraybackslash}p{#1}}

\definecolor{mn}{RGB}{255,127,0}
\definecolor{ts}{RGB}{0,0,255}
\definecolor{mr}{RGB}{190,0,80}

\newcommand{\vect}[1]{\mathbf{#1}}
\newcommand{\ModelPredictiveControl}{MPC}

\makeatletter
\newif\if@blind
\@blindfalse
\if@blind
\newcommand{\blind}[1]{\textbf{omitted for blind review}}
\newcommand{\citeblind}[1]{[omitted]}
\else
\newcommand{\blind}[1]{#1}
\newcommand{\citeblind}[1]{\cite{#1}}
\fi
\makeatother

\makeatletter
\newcounter{IEEE@bibentries}
\renewcommand\IEEEtriggeratref[1]{%
	\renewbibmacro{finentry}{%
		\stepcounter{IEEE@bibentries}%
		\ifthenelse{\equal{\value{IEEE@bibentries}}{#1}}
		{\finentry\@IEEEtriggercmd}
		{\finentry}%
	}%
}
\makeatother

\newcommand\copyrighttext{%
	\footnotesize%
	 \parbox[t]{.11\textwidth}{\copyright{} \the\year~IEEE.}%
	 \parbox[t]{.89\textwidth}{%
	 	Personal use of this material is permitted. Permission from IEEE must be obtained for all other uses, in any current or future media, including reprinting/republishing this material for advertising or promotional purposes, creating new collective works, for resale or redistribution to servers or lists, or reuse of any copyrighted component of this work in other works.%
 	 }%
}
\newcommand\copyrightnotice{%
	\tikzexternaldisable
	\begin{tikzpicture}[remember picture,overlay]
	\node[anchor=south,yshift=10pt] at (current page.south) {\parbox{\dimexpr\textwidth-\fboxsep-\fboxrule\relax}{\copyrighttext}};
	\end{tikzpicture}%
	\tikzexternalenable
}

\newcount\pgfplotstableuniqueentry
\def\pgfplotstableadduniquecol#1#2{%
	\pgfplotstablecreatecol[
	create col/assign/.code={%
		\getthisrow{#1}\entry\getnextrow{#1}\nextentry
		\ifx\entry\nextentry\relax
			\xdef\tempentry{\entry}
			\pgfkeyssetvalue{/pgfplots/table/create col/next content}{}
			\global\advance\pgfplotstableuniqueentry1\relax
		\else%
			\ifnum\the\pgfplotstableuniqueentry>0\relax
				\advance\pgfplotstableuniqueentry1
				\edef\temp{\noexpand
				\pgfkeyssetvalue{/pgfplots/table/create col/next content}{%
					\noexpand%
					\multirow{-\the\pgfplotstableuniqueentry}{*}{\tempentry}%
				}%
				}\temp
				\global\pgfplotstableuniqueentry=0
			\else%
				\pgfkeyslet{/pgfplots/table/create col/next content}{\entry}
			\fi%
		\fi%
		}
		]{u-#1}{#2}%
	}

\title{\LARGE \bf
Model Predictive Control Based Trajectory Generation for Autonomous Vehicles -- An Architectural Approach
}

\author{Marcus Nolte$^{1}$, Marcel Rose$^{1}$, Torben Stolte$^{1}$ and Markus Maurer$^{1}$
\thanks{$^{1}$All authors are with the Institute of Control Engineering,
        Technische Universit\"at Braunschweig, 38106 Braunschweig, Germany
        {\tt\small \{nolte,stolte,maurer\}@ifr.ing.tu-bs.de}}%
}

\begin{document}

\maketitle
\thispagestyle{empty}
\pagestyle{empty}

%
\begin{abstract}%
	Research in the field of automated driving has created promising results in the last years.
Some research groups have shown perception systems which are able to capture even complicated urban scenarios in great detail.
Yet, what is often missing are general-purpose path- or trajectory planners which are not designed for a specific purpose.
In this paper we look at path- and trajectory planning from an architectural point of view and show how model predictive frameworks can contribute to generalized path- and trajectory generation approaches for generating safe trajectories even in cases of system failures.

\end{abstract}%

\copyrightnotice
\section{Introduction}
\label{sec:introduction}
%
In recent years, the field of automated driving has gained much attention not only from research groups, but also from industry and even the public.
In public reception, the timespan until "autonomous" vehicles will be ready for market seems to be a matter of years rather than decades, also due to the high expectation raised by marketing strategies in industry.

The field of environment perception has recently seen impressive progress, also boosted by the developments in the field of machine learning.
This has increased the level of detail of the representation of the vehicle's environment as a basis for decision making and, eventually, for trajectory generation.
However, in most cases, the driver is still considered to be the ultimate fallback in case of system failures, so that current systems have to be considered \emph{partially automated} (level 2) according to the definition of SAE \cite{sae2014}.

When moving towards \emph{fully automated} (level 5) or even autonomous vehicles, safety requirements emerge as the system must be able to handle system failures without external intervention.
Thus, a detailed representation of the vehicle's environment alone is not sufficient. 
Furthermore the external representation must be combined with a detailed representation of the vehicle's internal states and its internal capabilities which are required to fulfill the vehicle's mission, as already demanded by Dickmanns \cite{dickmanns1987}, Maurer \cite{maurer2000}, Pellkofer \cite{Pellkofer2003}, Siedersberger \cite{Siedersberger2003}, and Reschka \cite{Reschka2017}.
By fusing the internal representation, e.g. about available actuators, with the external context extracted from sensor data, the vehicle becomes \emph{self-aware} and is enabled to take safe driving decisions at any time, even in case of system failures. 

Extracting the required information about the internal state and the external context is only one part of the realization of safe driving strategies.
Decision algorithms and eventually trajectory generation and control algorithms must be designed in such a way, that available information can be utilized efficiently at each level of abstraction.
As an example, a degradation of the steering system due to a failure has immediate impact on vehicle stability when turning so that control algorithms must be immediately reparameterized in order to keep the vehicle stable.
Additionally, a trajectory generation module -- which is running at a lower frequency -- must also plan trajectories which conform to the resulting actuator capabilities after degradation e.g. by generating trajectories with reduced curvature.
Eventually, decision algorithms must be aware of possible degradation effects in order to generate reachable target poses for the vehicle.

In this paper, we propose model predictive control (MPC) as a framework which can be utilized for reflecting performance degradation and mitigating the emerging degradation effects at the lower levels of a system architecture.
Corresponding to the above example, we will map decisions about constraints and optimization parameters to a functional system architecture and describe an actually implemented MPC framework for our full-by-wire research vehicles MAX and MOBILE.

This paper is structured as follows:
Section \ref{sec:related_work} gives a short overview over related work regarding functional system architectures, the demand for self-awareness, and recent approaches to {\ModelPredictiveControl} for automated road vehicles.
After this, Section \ref{sec:navigation} describes the utilized functional system architecture and how {\ModelPredictiveControl} can be mapped to it in combination with information about the internal vehicle state.
Section \ref{sec:mpc_rear_steer} presents a multi-stage {\ModelPredictiveControl} framework for autonomous vehicles, considering active rear steering.%
\section{Related Work}
\label{sec:related_work}
For the development of complex vehicle systems, the ISO~26262 standard requires architectural designs that feature \emph{modularity}, \emph{adequate level of granularity}, and \emph{simplicity} \cite[Part 3, 7.4.3.7]{iso_2011}.
Functional system architectures provide the possibility to fulfill these design requirements demanded by the ISO26262. 
In the field of automated driving, architectures which relate to the robotics \emph{sense, plan, act} scheme are utilized. 

One of the more recent approaches to a functional system architecture for level~3 to level~5 automated vehicles is presented by Matthaei and Maurer \cite{matthaei2015d}. 
Their contribution is based on earlier works of Rasmussen \cite{rasmussen1983}, Donges \cite{donges1999}, Dickmanns et al. \cite{dickmanns1994}, and Maurer \cite{maurer2000}.
A more detailed presentation of this contribution \cite{matthaei2015d} is presented in Section \ref{sec:navigation}.
A deeper review of existing architectural approaches for automated road vehicles has also been published by Matthaei and Maurer \cite{matthaei2015d}.
Ta{\c s} et al. \cite{tas2016a} compare architectures of selected automated vehicles in a unified graphical representation and also state the necessity for self- and performance monitoring.

With regard to the internal representation of capabilities of automated vehicles, first demands for representing the ability to accelerate, brake and turn have been stated by Dickmanns~ \cite{dickmanns1987}.
Based on Passino and Antsaklis \cite{passino1993f}, Maurer provides a first concept for a runtime representation of the performance of control algorithms in automated vehicles \cite{maurer2000}.
Further developing and implementing the idea, Pellkofer \cite{pellkofer2002} and Siedersberger \cite{Siedersberger2003} introduce the concept of skill networks.
A further differentiation between skills and abilities is made by Bergmiller \cite{bergmiller2015towards} and Reschka \cite{reschka2015ability}, the latter introducing the concept of ability and skill graphs for runtime monitoring and support of development processes.

With regard to {\ModelPredictiveControl}, several contributions have recently been made for trajectory generation and control in the field of automated driving.
An approach for lateral vehicle guidance, using MPC for trajectory planning and control is presented and evaluated in simulation by G\"otte et al. \cite{gotte2016}.
A similar approach utilizing more advanced collision avoidance constraints is described by Gutjahr and Werling \cite{gutjahr2016}.
Yi et al. \cite{yi2016} propose an approach to MPC-based trajectory planning for critical driving maneuvers. 
They introduce a quadratic MPC problem for considering friction limits in evasion maneuvers.

%
\section{MPC-Based Trajectory Generation in a Functional System Architecture}
\label{sec:navigation}
\begin{figure*}
	\centering
	\includegraphics[width=0.85\textwidth]{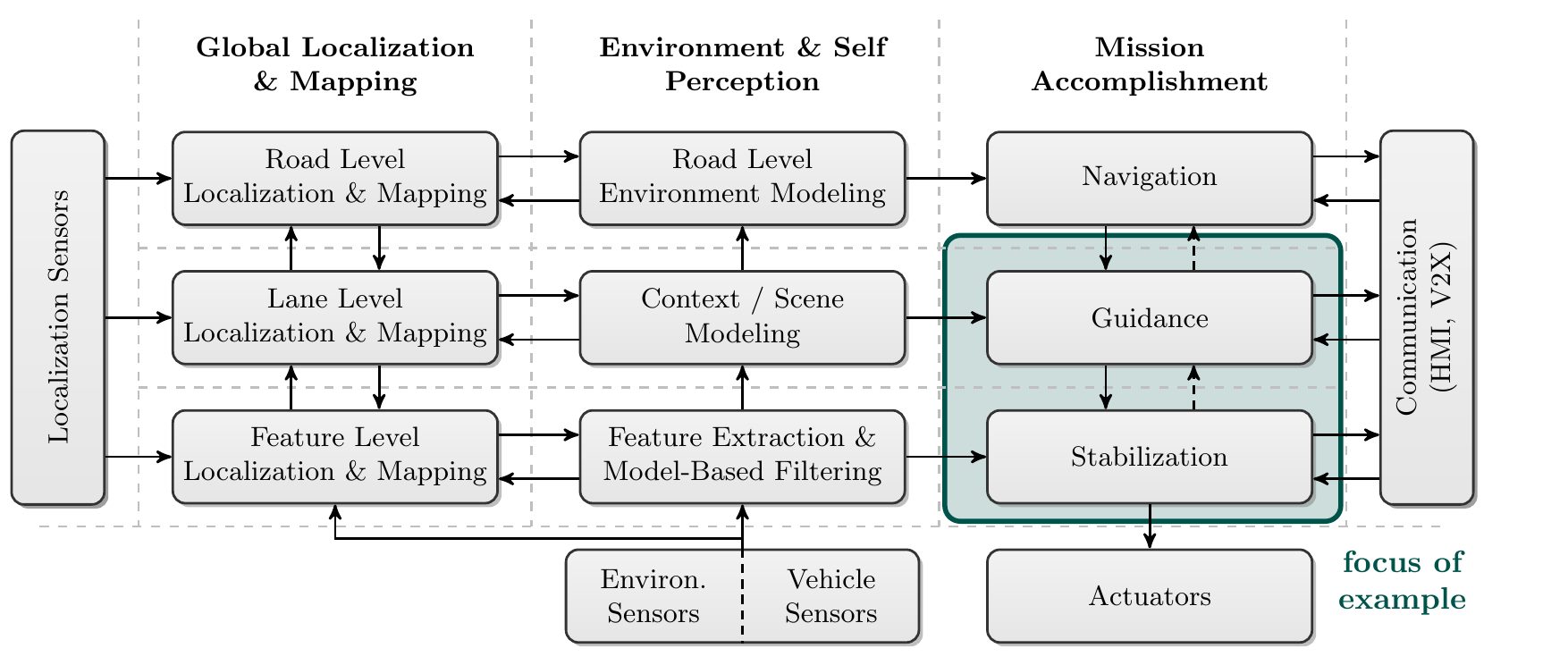}
	\caption{Functional system architecture based on \cite{matthaei2015d} and \cite{ulbrich_2015}: This paper focuses on \emph{Guidance} and \emph{Stabilization} in \emph{Mission Accomplishment}.
	}
	\label{fig:fusysarch}
	\vspace{-1em}
\end{figure*}
Model predictive control aims at generating optimal control inputs for a given plant.
Thus, a weighted cost function as well as optimization constraints are required in order to fully describe the related optimization problem.
In case of trajectory generation for automated vehicles, optimization constraints are e.g.~based on the vehicle's environment, as the generated trajectory must always be collision free.
Moreover, the internal state of the vehicle can introduce additional constraints, e.g.~ by incorporating actuator limitations such as the maximum angular steering change rate.
Weights in the cost function emerge from 
additional mission parameters (e.g. comfort, actuator wear or energy consumption).

In the following subsections, we will describe how these constraints and weights are generated at different levels of the functional system architecture.
For this, we will relate to the architectural approach proposed by Matthaei and Maurer~\cite{matthaei2015d}.
The resulting architecture is displayed in Fig.~\ref{fig:fusysarch}.
The authors consider the sense, plan, act scheme and derive a three-layer architecture operating at road level (\emph{navigation}), lane level (\emph{guidance}), and at a quasi-continuous level (\emph{stabilization}).
The three levels are divided into three columns each: \emph{global localization \& mapping}, \emph{environment and self-perception}, and \emph{mission accomplishment}.

\begin{figure}
	\centering
	\includegraphics[width=0.95\columnwidth]{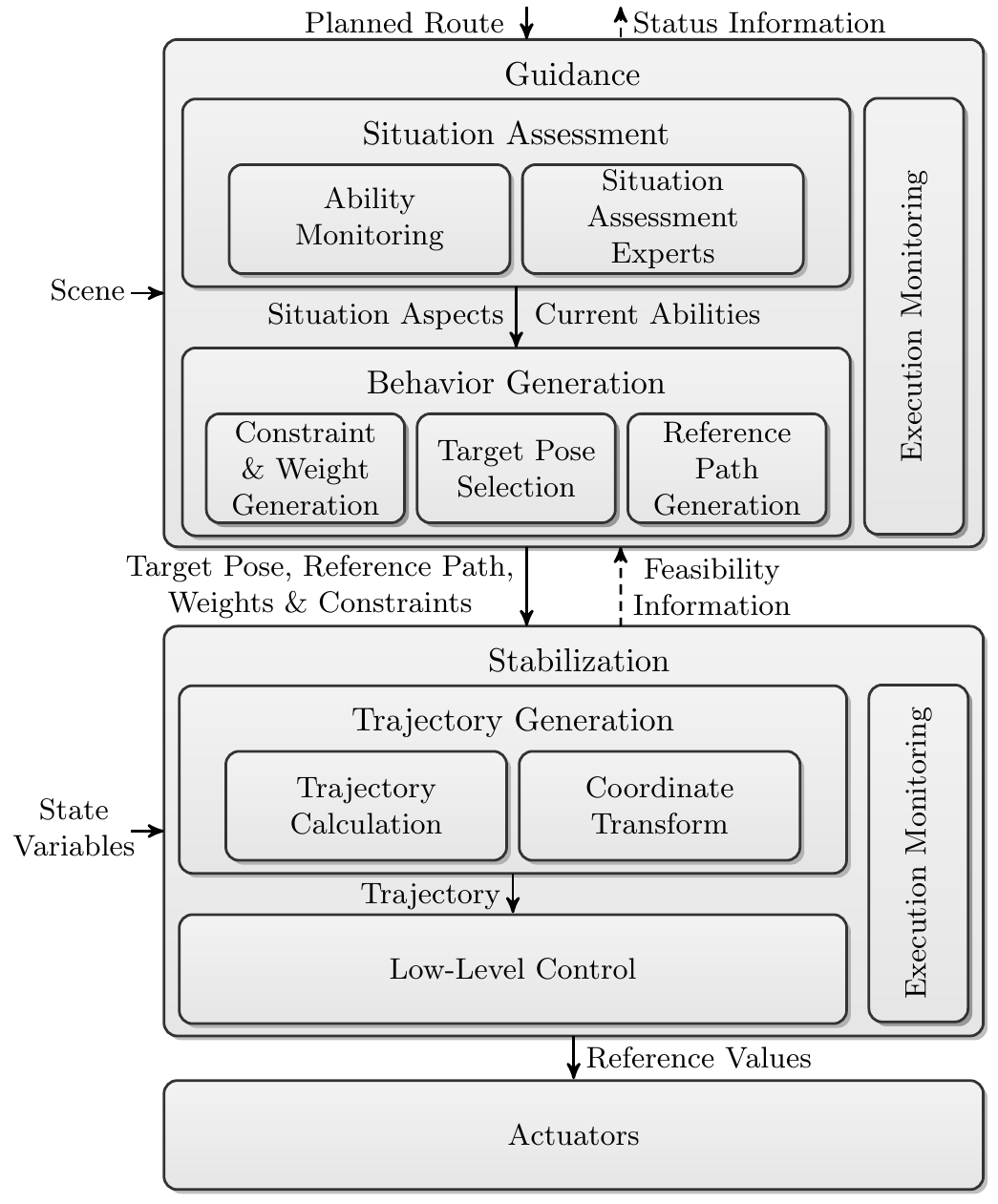}
	\caption{Detailed view at Guidance and Stabilization}
	\label{fig:missionArchitecture}
	\vspace{-0.58em}
\end{figure}

Fig.~\ref{fig:missionArchitecture} depicts a closer view at the guidance and stabilization level for mission accomplishment, which is the focus in the following. 
However, a short description for the navigation level is given, as the planned route provides the basis for selecting a target pose at lower architecture levels.
\subsection{Navigation}
At the navigation level (cf.~Fig. \ref{fig:fusysarch}), the most recent available information about the road network is provided from the environment perception column supported by a-priori external data (maps) according to Matthaei and Maurer \cite{matthaei2015d}.
Applied algorithms for route planning are typical graph-based approaches such as Dijkstra or A* derivatives.
Mission goals as well as possible information from self-representation modules can be represented as edge weights or costs in the respective graph.
Thus, only routes will be planned, which conform with mission goals and the vehicle's abilities (e.g. sloped roads could be avoided if the braking system can only provide reduced functionality).
Typical cycle times at this level are several ten seconds or even minutes. 
Finally, the planned route will then be input for the algorithms at guidance level.

\subsection{Guidance}
\label{subsec:planning}
At the guidance level, all tactical decisions which concern the immediate surroundings of the vehicle are made. 
The goal at the guidance level is twofold:
On the one hand, a target pose as input for trajectory generation shall be generated.
This target pose must be reachable and it must conform to all (i.e. legal) given constraints.  
As an example, in ordinary driving scenarios, the target pose will be located either in the ego-lane or in a neighboring lane in case a lane change is desired.
On the other hand, a unique reference path which leads to that target pose must be calculated to facilitate the application of {\ModelPredictiveControl}.
In a structured environment, this typically is the lane center line.
By this, a trajectory can be generated along that reference path, making the trajectory a function of lateral distance from the reference.
The reference path does not need to be smooth, as smoothness of the resulting trajectory will be ensured by the trajectory generation modules at the stabilization level.
The reference generation (cf.~Fig.~\ref{fig:missionArchitecture}) can be seen as a coarse path planning in the immediate static environment of the vehicle with a horizon of up to several hundred meters.
Thus, graph based approaches can be applied at this level of granularity, e.g. based on visibility graphs, cf.~\cite{VisiLibity:08}.
 
In order to generate a target pose and a reference path, the scene (cf.~\cite{ulbrich_2015} for definition), which is the result from context and scene modeling, is combined with the planned route, mission aspects (e.g. drivable area or cooperation aspects), as well as constraints and parameters generated from self-representation (cf.~Fig.\ref{fig:missionArchitecture}, Ability Monitoring).

In terms of \ModelPredictiveControl, the static as well as the dynamic environment will be transformed to spatial constraints in the vehicle's coordinate frame.
This does not only include rigid objects, but also those areas which must not be crossed due to legal regulations (e.g. solid lines).
However, all decisions about constraints at this level are of tactical implication.
Actuation constraints can be imposed at this level, if they are part of an optimization process.
Furthermore, actuator degradation can be considered at this level. 
In case degrading actuator performance has been detected, an immediate reaction to stabilize the vehicle must be performed at the stabilization level. 
However, the system must also plan degraded trajectories, which account for degraded actuation abilities (e.g.\ constraints on maximal available steering angles).

Additionally, tactical decisions about optimization costs will be made here.
For instance, if the vehicle is driving with passengers, optimization weights for limiting jerk and yaw rate will be increased, in order to increase passenger comfort.
If the vehicle is driving without passengers, a higher maneuverability might be desirable, such that those weights will be lowered.
 
\begin{figure}[h!]
	\centering
	\scalebox{0.9}{%
	\input{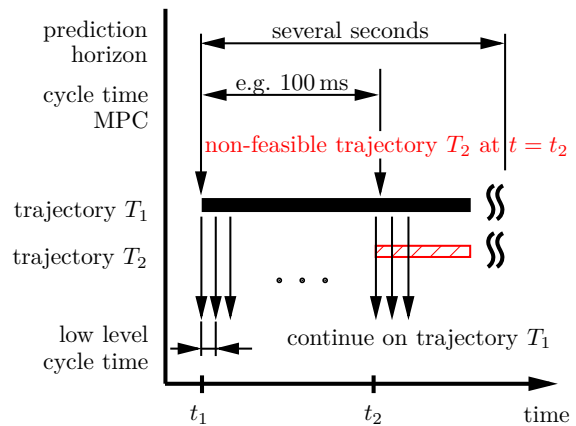}
	}
	\caption{Sequential display of a low-level controller using a fallback trajectory (e.g.~ for braking the vehicle) in case of infeasible optimization problem at time $t_2$.}
	\label{fig:fallbackTrajectory}
	\vspace{-0.5em}
\end{figure}

Eventually, the guidance level must ensure that at least one trajectory is available which can be used for safely stopping the vehicle. 
This can be reached by making use of the least valid trajectory in case of infeasible optimization problems as visualized in Fig.~\ref{fig:fallbackTrajectory}.
In emergency situations, the above requirement can also demand a re-parameterization of optimization constraints.
To clarify this, consider the following situation:
An imminent crash can only be avoided by performing an evasion maneuver.
The target pose is located in a valid place, but it is not reachable because a solid line would need to be crossed.
Under normal conditions, legal constraints would prohibit this maneuver, which would be reflected in the spatial optimization constraints.
The resulting optimization problem would be over-constraint, without possessing a feasible solution.
This is why the guidance level needs additional feasibility information from the stabilization level.

If a non-feasible optimization problem is detected, several strategies could be applied.
Providing several levels of drivable area could be considered:
Normally, solid lines must not be crossed, as this is prohibited by law.
In emergency situations, this constraint could be lifted at the tactical level if no feasible trajectory is found.
In the following iteration, with changed spatial constraints, the subordinate MPC framework could then find a feasible solution.

\subsection{Stabilization}
\label{subsec:stab}
Taking the reference path and the spatial constraints from the guidance level as an input, Matthaei and Maurer~\cite{matthaei2015d} divide the stabilization level into two sub levels.
These levels are responsible for trajectory generation and for trajectory following control.
Typical cycle times for the generation layer are multiple ten to hundred milliseconds and several milliseconds for the control layer.

With regard to {\ModelPredictiveControl}, multiple approaches can be considered:
On the one hand, the MPC framework can be utilized for model based trajectory generation only in combination with a subordinate lateral and longitudinal controller as proposed by Falcone et al. \cite{falcone2008}.
This is a feasible approach, e.g. if the model reflects basic vehicle properties, but lacks accuracy to perform actual model based control, particularly if appropriate disturbance models are lacking.
In addition, the problem of long cycle times due to a slowly solvable optimization problem can be circumvented using an additional controller:
While the MPC framework calculates a new trajectory, the controller(s) can always work on the last valid reference values.
On the other hand, if the MPC algorithm is real-time capable and can be provided with highly accurate vehicle models, the additional controller can be omitted.

Considering data from self-perception modules, actuation constraints are modified in order to adopt to the actual system state.
For instance, in case of a mechanical brake with degraded performance, trajectories with reduced decelerations will be planned.  
 
From a functional safety perspective, MPC as framework at the stabilization level allows exploiting functional redundancies to compensate for degradation of the actuation system 
due to failures, wear, or insufficient energy supply. 
Perceived degradation is mapped to the constraints related to the affected actuators. 
In case of safety relevant degradation, the controller will access all other available actuators in order to follow the planned trajectory. 
For instance, a free-running wheel due to a failure of a mechanical brake leads to reduced deceleration capabilities as well as to undesired yaw torque when decelerating.
In this case, the controller compensates for reduced deceleration capabilities by utilizing the drive train to reduce speed
as well as other available actuators in order to compensate for the yaw moment induced by the degraded brake (e.g. by additional steering angle, differential braking, or torque vectoring).




Output data of the stabilization level consists of direct physical reference values (e.g. steering angle, motor torque) for the respective actuators.%
\section{MPC Framework for an Automated Vehicle with Active Rear Steering}
\label{sec:mpc_rear_steer}
The following section will present an actual implementation of an {\ModelPredictiveControl} framework for our electric, over-actuated, full-by-wire experimental vehicles.
Implementation will be described related to the architectural concerns presented in Section \ref{sec:navigation}.

\subsection{Experimental Vehicles \& Problem Statement}
\label{sec:vehicles}
\begin{figure}[h]
	\centering
	\includegraphics[width=.75\columnwidth]{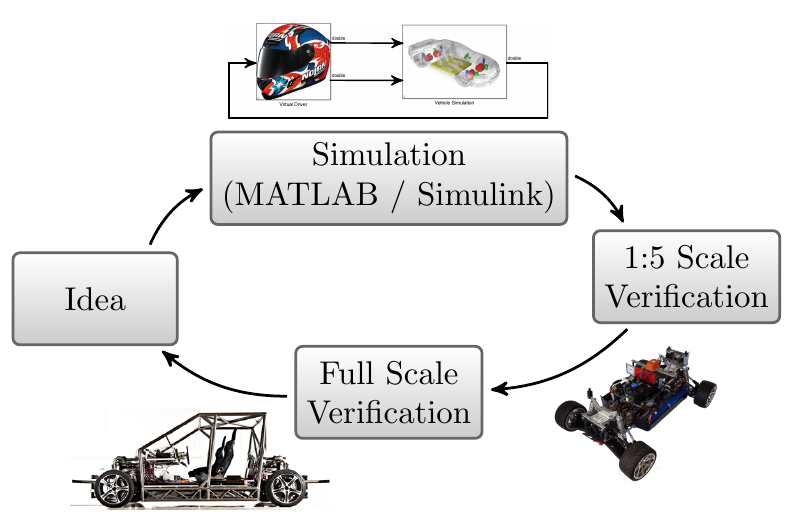}
	\caption{Tool-chain with vehicles MOBILE (left) and MAX (right)}
	\label{fig:toolchain}
\end{figure}

The proposed MPC framework for trajectory planning was implemented for our research vehicles MOBILE and MAX. 
Both vehicles are part of a rapid-prototyping tool chain (Fig.~\ref{fig:toolchain}) developed at the Institute of Control Engineering.
This MATLAB/Simulink tool chain was implemented to perform early tests of vehicle control algorithms on a 1:5 model vehicle (MAX) in order to allow for safe testing, before moving to full-scale validation (MOBILE).

Both vehicles are equipped with the same ECUs and feature a FlexRay communication backbone as well as CAN bus connections for communication with sensors and actuators.
Regarding the actuator topology, both vehicles are equipped with four close-to-wheel drives and four individually steerable wheels.
For further details about the drive train and network topologies, refer to~\cite{bergmiller2015towards}.
Regarding the actuator topology, model based control strategies can serve as a tool for efficient vehicle dynamics control e.g. by utilizing torque vectoring or force allocation approaches.
Moreover, functional redundancies can be synthesized e.g. by utilizing single wheel steering and recuperation to generate additional brake force, if the brake system is only available with degraded functionality.
In addition, the full-scale vehicle MOBILE is equipped with a set of LiDAR Sensors ($2\times$~Velodyne VLP-16 PUCK and a Velodyne HDL-32) for basic environment perception.

For exploiting the vehicle's capabilities even under ordinary driving conditions, coordination between actuators needs to be performed as the driver's input capabilities are limited. 
E.g. for steering, the driver can only command a single steering angle $\delta$ at the steering wheel, while the wheels are steered with a front ($\delta_f$) and a rear steering angle ($\delta_r$). 
By coordinating rear and front steering, the turn radius of the vehicle can be modified depending on the actual driving scenario. 
On the one hand, a smaller turn radius can be realized by counter steering (e.g.~$\delta_f = -\delta_r$), thus creating higher yaw rates and making the vehicle more agile, for instance when parking.
On the other hand, a larger turn radius, or even parallel steering (e.g.~$\delta_f = \delta_r$) can be used at higher velocities to increase driving comfort for the passengers.

Our proposed MPC framework shall be able to perform the above mentioned coordination between rear- and front axle steering, as well as the enforcement of constraints derived from self-representation as described in Section~\ref{sec:navigation} in a static and unstructured vehicle environment.
For this purpose, we focus on the example of lateral planning, such that constraints regarding the ability to steer the vehicle will be reflected in terms of available steering angle and steering angle rate at the front and rear axle.
Additionally, as this paper shall focus on the trajectory planning framework.
We assume that our system is provided with information about the static environment and that information about the maximal available steering angles can be derived from self-representation.
As a first step, the implementation is evaluated in simulation. 
The vehicle model used for simulation is a continuous time non-linear double track model, while the MPC strategy is based on an LPV single track model (equations \ref{eq:AMatrix}, \ref{eq:BMatrix}, and \ref{eq:CMatrix}).

\subsection{Implementation}
In conformance with the proposed functional system architecture, we propose a two staged approach:
At the \emph{guidance level} (cf. Section~\ref{subsec:planning}) an unambiguous reference path is extracted from the static environment, taking into account all static obstacles.
Spatial constraints which are required for performing actual {\ModelPredictiveControl} at the \emph{stabilization level} (cf.\ Section~\ref{subsec:stab}) are transformed from a vehicle relative cartesian frame to a Frenet frame along the reference.
Actual trajectory planning is performed at the stabilization level in Frenet space.

\subsubsection{Guidance Level}
For reference path generation we assume, that the static environment is represented by an occupancy grid based.
By extracting occupied as well as free space from this grid based representation we arrive at a polygonal based description of impassable areas around the vehicles.

In a first step, a visibility graph is constructed from the polygonal environment utilizing \cite{VisiLibity:08}.
This visibility graph contains all polygon vertices as well as the start and the goal point as nodes.
The graph's edges are determined by connecting all nodes which are in line of sight, to prevent edges going through an obstacle. Weights for this edges are given by euclidean distance $s$ between two vertices.

\begin{figure}[t]
	\centering
	\scalebox{0.8}{%
	\input{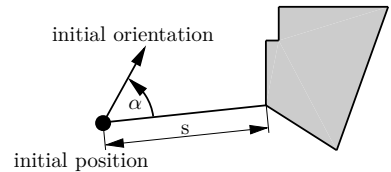}
}
	\caption{Penalty for vehicle's orientation relative to first node starting from initial position node}
	\label{fig:headingCost}
	\vspace{-1em}
\end{figure}

\begin{figure}[h]
	\centering
	\scalebox{0.85}{%
	\input{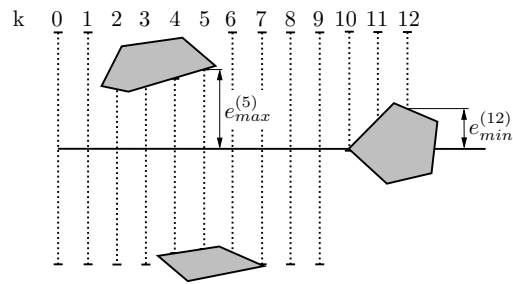}%
	}
	\caption{Definition of constraints in relation to reference path}
	\label{fig:constraints}
\end{figure}

In order to determine the shortest path through this visibility graph, we use an A* approach~\cite{VisiLibity:08}.
However, in order to not only consider the initial and terminal position of the vehicle, but also the initial and terminal orientation, additional edge weights have been defined which are added to the initial weight $s$ in the A* implementation.
These weights $w_{\mathit{start, end}}$ penalize the angle $\alpha_{\mathit{start, end}}$ between the vehicle's longitudinal axis and the edges connected to the first/last node in the visibility graph (cf.~Fig.~\ref{fig:headingCost}).
The total cost $C$ for traveling along the respective edges is then modified to
\begin{equation}
	C = s + w_{\mathit{start, end}} \cdot |\alpha_{\mathit{start, end}}|.
\end{equation}
The resulting shortest path will be utilized as a reference path and is represented by a polyline through the static environment.
In order to derive the spatial constraints required for the MPC framework, a maximal and a minimal deviation $e_{\mathit{max, min}}$ from the reference path are defined and superimposed with the orthogonal distances of the given polygon edges from the polyline (cf.~Fig~\ref{fig:constraints}).

This way, linear spatial constraints for the static environment are generated for each sampling interval $k$ in the MPC framework.

Additional constraints to those already mentioned (Section \ref{sec:vehicles}) which are generated at this level are constraints for maximal side slip angle and maximal yaw rate in order to influence overall driving behavior.
In our example, strategic decisions about optimization weights relate to weighting front and rear axle steering by formulating penalizing either high yaw rates (for more side slip angle in the resulting trajectory) or high side slip angle (for more yaw rate in the resulting trajectory).

\subsubsection{Stabilization Level}
\begin{figure}[h]
	\centering
	\scalebox{0.75}{%
	\input{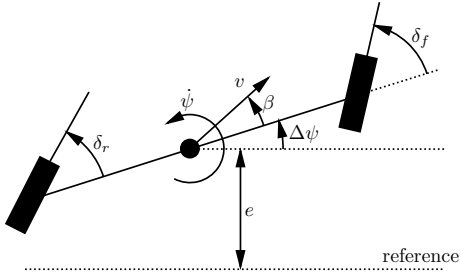}
	}
	\vspace{-0.8em}
	\caption{Single track model}
	\label{fig:singleTrackModel}
\end{figure}

The actual MPC for trajectory generation is performed at the stabilization level implemented with the MATLAB MPC Toolbox.
As a first step, we assume that an extended linearized single track model with active rear steering is suitable to generate a basic collision free lateral trajectory.
I.e. we assume linear tire dynamics as well as small side slip and heading angles.
For the required states we follow an approach presented by Gutjahr and Werling \cite{gutjahr2016}.
Apart from the side slip angle $\beta$ and the vehicle's yaw rate $\dot \psi$, we introduce the heading difference to the reference frame $\Delta\psi$ and lateral deviation in Frenet coordinates $e$ in order to describe the motion of the vehicle.
Thus, our state space vector $\vect x$ is:
\begin{equation}
	\vect{x} = \left[\beta \quad \dot\psi \quad \Delta\psi \quad e\right]^T
\end{equation}
Inputs to the model are front and rear steering angles $\delta_f, \delta_r$, as well as a disturbance term $\Delta \psi_{\mathit{ref}}$ to model the trajectory's curvature in the world frame (cf.~Fig.~\ref{fig:singleTrackModel}).

The resulting input vector $\vect{u}$  is 
\begin{equation}
	\vect{u} = \left[\delta_f \quad \delta_r \quad \Delta\psi_{\mathit{ref}}\right]^T.
\end{equation}
By considering a linear parameter variant (LPV) single track model with vehicle mass $m$ and velocity $v$ ($v$ as a time-variant parameter), the front and rear cornering stiffness $c_{\alpha,f}$ and $c_{\alpha,r}$, front and rear distance from the vehicle's center of gravity $l_f$ and $l_r$ as well as the moment of inertia around the vehicle's $z$-axis $J_z$, the resulting system matrix $\vect{A}(v)$ and input matrix $\vect{B}(v)$ are given by
\begin{equation}
	\vect{A}(v) = 
	\begin{bmatrix}
-\frac{c_{\alpha,f}+c_{\alpha,r}}{m \cdot v} & 
\frac{c_{\alpha,r} \cdot l_{r}-c_{\alpha,f} \cdot l_{f}}{m \cdot v^2} -1& 0 & 0 \\ 
-\frac{c_{\alpha,f} \cdot l_{f}-c_{\alpha,r} \cdot l_{r}}{J{z}} & 
-\frac{c_{\alpha,f} \cdot l_{f}^2+c_{\alpha,r} \cdot l_{r}^2}{J{z} \cdot v} & 0 & 0\\ 
0 & 1 & 0 & 0 \\
v & 0 & v & 0
\end{bmatrix}
\label{eq:AMatrix}
\end{equation}
and
\begin{equation}
	\vect{B}(v) = 
	\begin{bmatrix}
		\frac{c_{\alpha,f}}{m \cdot f} & \frac{c_{\alpha,r}}{m \cdot v} & 0 \\
		\frac{c_{\alpha,f} \cdot l_f}{J_z} & \frac{c_{\alpha,r} \cdot l_r}{J_z} & 0\\ 
		0 & 0 & -1\\ 
		0 & 0 & 0
	\end{bmatrix}.
\label{eq:BMatrix}	
\end{equation}
In order to check for collisions during MPC calculation, it is not sufficient to impose optimization constraints on the center of gravity of the vehicle, as this does not consider the vehicle's orientation.
Hence, we do not only consider the lateral distance of the vehicle's center of gravity $e$, but also the lateral distances of the vehicle's front $e_f$ and the vehicle's rear $e_r$ from the reference path, which are given by
\begin{equation}
e_{\mathit{f,r}} = e \pm l_{\mathit{f,r}} \cdot \sin(\Delta\psi).
\end{equation}

By introducing the linearized distances in the measurement matrix $\vect{C}$, all required dimensions can be constraint during optimization:
\begin{equation}
	\vect{C} = 
\begin{bmatrix}
	1 & 0 & 0 & 0 \\
	0 & 1 & 0 & 0 \\
	0 & 0 & 1 & 0 \\
	0 & 0 & 0 & 1 \\
	0 & 0 & l_{f} & 1 \\
	0 & 0 & -l_{r} & 1 \\
\end{bmatrix} 
\label{eq:CMatrix}
\end{equation}

For applying the {\ModelPredictiveControl} scheme to the defined system, we formulate a cost function as follows:
The total cost $J(\vect{z_k})$ for a given series of $p$ future system inputs $\vect u$ at the current control step $k$
\begin{equation}
	\vect{z_k}^T = \left[ \vect{u}(k|k)^T \ \vect{u}(k+1|k)^T \dots \ \vect{u}(k+p-1|k)^T\right]
\end{equation}

can be described as the sum of costs of depending on the following properties ($n_y = \dim(\vect y)$, $w^y_{i,j}$ weight for output $y_j$ at prediction step $i$): \\
1) the weighted deviation of measured outputs $\vect y$ from their given reference $\vect r$  
\begin{equation}
	J_y(\vect z_k) = \sum_{j=1}^{n_y} \sum_{i=1}^{p} \left(w^y_{i,j}\left[r_j(k+i|k) - y_j(k+i|k)\right]\right)^2,
\end{equation}

2) the weighted deviation of calculated inputs $\vect u$ from a targeted series of input vectors $\vect{u_t}$
\begin{equation}
J_u(\vect z_k) = \sum_{j=1}^{n_u}\sum_{i=1}^{p} \left(w^u_{i,j} \left[u_j(k+i|k) - u_{\mathit{t}, j}(k+i|k)\right]\right)^2,
\end{equation}

3) the weighted difference between input vectors in subsequent time steps
\begin{equation}
J_{\Delta u}(\vect z_k) = \sum_{j=1}^{n_{\Delta u}}\sum_{i=1}^{p} \left(w^{\Delta u}_{i,j} \left[\vect u_j(k+i|k) - u_j(k+i-1|k)\right]\right)^2.
\end{equation}

Combined, this yields the complete cost function
\begin{equation}
	J(\vect z_k) = J_y(\vect z_k) + J_u(\vect z_k) + J_{\Delta u}(\vect z_k).
\end{equation}

Please note, that slack variables are omitted for brevity (please refer to \cite{matlab} for further information).

In the actual implementation, we utilize the output weights $w^y_{\beta}, w^y_{\dot\psi}, w^y_{\Delta\psi}$, and $w^y_{e}$ for tracking the corresponding output variables.
Absolute front and rear steering angles and their corresponding rates are minimized using the weights $w^u_{\delta_f}$ and $w^u_{\delta_r}$ as well as $w^{\Delta u}_{\dot\delta_f}$ and $w^{\Delta u}_{ \dot\delta_r}$

The full optimization problem, including all constraints for states and inputs, results in solving
\begin{equation*}
	\min(J(\vect z_k)) \quad \text{subject to}
\end{equation*}
\vspace{-0.8em}
\begin{eqnarray}
	\begin{split}
		\beta_{\mathit{min}}(i) &\leq& \beta(k+i|k) &\leq& \beta_{\mathit{max}}(i)\\
		\dot\psi_{\mathit{min}}(i) &\leq& \dot\psi(k+i|k) &\leq& \dot\psi_{\mathit{max}}(i)\\
		e_{\mathit{min}}(i) &\leq& e(k+i|k) &\leq& e_{\mathit{max}}(i)\\
		e_{f,\mathit{min}}(i) &\leq& e_f(k+i|k) &\leq& e_{f,\mathit{max}}(i)\\
		e_{r,\mathit{min}}(i) &\leq& e_r(k+i|k) &\leq& e_{r,\mathit{max}}(i)\\		
		\delta_{f,\mathit{min}}(i) &\leq& \delta_f(k+i|k) &\leq& \delta_{f,\mathit{max}}(i)\\
		\delta_{r,\mathit{min}}(i) &\leq& \delta_r(k+i|k) &\leq& \delta_{r,\mathit{max}}(i)\\
		\dot\delta_{f,\mathit{min}}(i) &\leq& \dot\delta_f(k+i|k) &\leq& \dot\delta_{f,\mathit{max}}(i)\\
		\dot\delta_{r,\mathit{min}}(i) &\leq& \dot\delta_r(k+i|k) &\leq& \dot\delta_{r,\mathit{max}}(i)\\		
	\end{split}.
\end{eqnarray}
As already stated, weights will be derived from the tactical level, constraints can originate from either the tactical or the stabilization level.
Yet, as the stabilization level is ideally immediately aware of sudden actuator degradations (e.g. by residual fault diagnosis techniques as proposed by Ding \cite{ding2013}), while it can take several hundred milliseconds until the planning level is aware of such degradations, constraints imposed at the stabilization level need to possess higher priority compared to constraints imposed at the planning level.
Regarding MPC there are two drawbacks, which must be considered: First, if MPC is used for actual control, the performance is highly dependent of accurate models for predicting model states. 
In addition, MPC can only adapt to new parameters between two optimization cycles.
As this can take several ten to hundred milliseconds, this might be to slow for sudden actuator failures.
Thus we require an underlying low level controller to be present for performing actual trajectory following control.

\subsection{Results}
\begin{figure}[b!]
	\vspace{-0.8em}
	\centering
	\subfloat[Trajectory and vehicle contour]{
%
%
\definecolor{mycolor1}{rgb}{0.00000,0.44700,0.74100}%
\definecolor{mycolor2}{rgb}{1.00000,1.00000,0.00000}%
\definecolor{mycolor3}{rgb}{0.85000,0.32500,0.09800}%
\begin{tikzpicture}

\begin{axis}[%
width=2.521in,
height=0.898in,
at={(0.542in,0.579in)},
scale only axis,
xmin=-0.0055,
xmax=5.5055,
xlabel={x in m},
ymin=-0.0017,
ymax=1.7017,
ylabel={y in m},
axis background/.style={fill=white},
]

\addplot[area legend,solid,line width=0.5pt,draw=black,fill=white,forget plot]
table[row sep=crcr] {%
x	y\\
0	0\\
5.5	0\\
5.5	1.7\\
0	1.7\\
}--cycle;

\addplot[area legend,solid,line width=0.5pt,draw=black,fill=white!80!black,forget plot]
table[row sep=crcr] {%
x	y\\
0.5	0.15\\
0.5	0.5\\
1.6	0.5\\
1.6	0.15\\
}--cycle;

\addplot[area legend,solid,line width=0.5pt,draw=black,fill=white!80!black,forget plot]
table[row sep=crcr] {%
x	y\\
1.8	0.15\\
1.8	0.5\\
2.9	0.5\\
2.9	0.15\\
}--cycle;

\addplot[area legend,solid,line width=0.5pt,draw=black,fill=white!80!black,forget plot]
table[row sep=crcr] {%
x	y\\
4.2	0.15\\
4.2	0.5\\
5	0.5\\
5	0.15\\
}--cycle;
\addplot [color=black,mark size=1.5pt,only marks,mark=*,mark options={solid,fill=black},forget plot]
  table[row sep=crcr]{%
1.3	1.1\\
};
\addplot [color=black,mark size=1.5pt,only marks,mark=*,mark options={solid,fill=black},forget plot]
  table[row sep=crcr]{%
3.7	0.35\\
};
\addplot [color=black!50!green,solid,line width=1.0pt,forget plot]
  table[row sep=crcr]{%
1.29474144342774	1.07826656211339\\
1.35485613544305	1.08206863199823\\
1.41503103016899	1.08553249494174\\
1.47530821129786	1.08835160168494\\
1.53570416901365	1.09037589822183\\
1.59621944041044	1.0914977363291\\
1.65684294763357	1.09163205523008\\
1.71755657658435	1.09072497025153\\
1.7783378587924	1.08874439196433\\
1.83915944847891	1.08565742416078\\
1.89999231519352	1.08144571968281\\
1.96080613388134	1.07609777814051\\
2.02156989917825	1.06960700188799\\
2.08225349728092	1.06197546264625\\
2.14282729734159	1.05320802654875\\
2.20326200834029	1.04331018043944\\
2.26352657323397	1.03227960939603\\
2.32358938820549	1.0201123096915\\
2.38341862413757	1.00680378158197\\
2.44298254397125	0.992349916928214\\
2.50225164409331	0.976753171333813\\
2.56119625413834	0.96001426432246\\
2.61978236732327	0.942121236394593\\
2.67797586286721	0.923063245334448\\
2.73574036740808	0.902825214975937\\
2.79303745337574	0.881389584116911\\
2.84982712787022	0.858739271918245\\
2.90606522337049	0.834858309611541\\
2.96168725180318	0.809737594024196\\
3.01660518376649	0.783384032507457\\
3.07075481456157	0.755810443807524\\
3.12410698546455	0.72701501923454\\
3.17665206995642	0.696986542965148\\
3.22839521116775	0.665713964765454\\
3.27936292359361	0.633200170463765\\
3.3296156466501	0.59948589066982\\
3.37931607790518	0.564792046466038\\
3.42879128959711	0.529662489654149\\
3.47842515937589	0.494769467423368\\
3.52857219660825	0.460762408583035\\
3.57953791844106	0.428249760054833\\
3.63154003526199	0.397720881009674\\
3.67941406029409	0.372473489193282\\
3.72298319962347	0.352247627456127\\
};
\addplot [color=black,solid,forget plot]
  table[row sep=crcr]{%
1.65631796678973	1.30162147643912\\
1.6816583974051	0.902424955191471\\
0.983064485221704	0.858079201614573\\
0.957724054606334	1.25727572286223\\
1.65631796678973	1.30162147643912\\
};
\addplot [color=black,solid,forget plot]
  table[row sep=crcr]{%
1.96751541843118	1.29829303778834\\
1.97479911103867	0.898359358508885\\
1.27491517229963	0.885612896445789\\
1.26763147969214	1.28554657572524\\
1.96751541843118	1.29829303778834\\
};
\addplot [color=black,solid,forget plot]
  table[row sep=crcr]{%
2.28420421699081	1.26312184501938\\
2.2649071465781	0.863587587322391\\
1.56572219560836	0.897357460544639\\
1.58501926602107	1.29689171824163\\
2.28420421699081	1.26312184501938\\
};
\addplot [color=black,solid,forget plot]
  table[row sep=crcr]{%
2.59865516644933	1.19916027033535\\
2.55295861744262	0.801779061155951\\
1.85754150137868	0.88174802191769\\
1.90323805038539	1.27912923109708\\
2.59865516644933	1.19916027033535\\
};
\addplot [color=black,solid,forget plot]
  table[row sep=crcr]{%
2.90639125988689	1.10826807902956\\
2.83661169942115	0.714401587764803\\
2.14734533970783	0.836515818579853\\
2.21712490017357	1.23038230984461\\
2.90639125988689	1.10826807902956\\
};
\addplot [color=black,solid,forget plot]
  table[row sep=crcr]{%
3.20354925852429	0.991412571868025\\
3.11309945230025	0.601773209695728\\
2.43123056849873	0.760060370587792\\
2.52168037472277	1.14969973276009\\
3.20354925852429	0.991412571868025\\
};
\addplot [color=black,solid,forget plot]
  table[row sep=crcr]{%
3.48481130102333	0.851634376198875\\
3.3809999290746	0.465340189012317\\
2.70498510149813	0.647010089922583\\
2.80879647344685	1.03330427710914\\
3.48481130102333	0.851634376198875\\
};
\addplot [color=black,solid,forget plot]
  table[row sep=crcr]{%
3.74183742897695	0.702918004518819\\
3.64515687252685	0.314777733476893\\
2.96591139820348	0.483968707264559\\
3.06259195465358	0.872108978306486\\
3.74183742897695	0.702918004518819\\
};
\addplot [color=black,solid,forget plot]
  table[row sep=crcr]{%
3.97451740031109	0.585145295050828\\
3.9298722609328	0.187644588724546\\
3.23424602486181	0.265773582636551\\
3.2788911642401	0.663274288962833\\
3.97451740031109	0.585145295050828\\
};
\addplot [color=black,solid,forget plot]
  table[row sep=crcr]{%
-0.577554513380057	-1.04652752770607\\
-0.803297174061062	-1.37673996107081\\
-1.38116893244935	-0.981690304879052\\
-1.15542627176835	-0.651477871514315\\
-0.577554513380057	-1.04652752770607\\
};
\addplot [color=black,solid,forget plot]
  table[row sep=crcr]{%
4.10391918642217	0.540695959796045\\
4.09169701843375	0.140882730137917\\
3.39202386653203	0.162271524117656\\
3.40424603452045	0.562084753775783\\
4.10391918642217	0.540695959796045\\
};
\end{axis}
\end{tikzpicture}%
		\label{fig:parkingSzenario}}
	\vspace{-1em}
	\hfil
	\subfloat[Steering angles]{	
%
%
%
\begin{tikzpicture}

\begin{axis}[%
width=2.521in,
height=1.05in,
at={(0.537in,0.441in)},
scale only axis,
xmin=0,
xmax=50,
xlabel={prediction step},
xmajorgrids,
ymin=-0.6,
ymax=0.1,
ylabel={steering angle in rad},
ymajorgrids,
axis background/.style={fill=white},
legend style={at={(0.03,0.03)},anchor=south west,legend cell align=left,align=left,draw=white!15!black}
]
\addplot [color=blue,solid]
  table[row sep=crcr]{%
1	2.54429399941181e-17\\
2	-0.0263260803307667\\
3	-0.0467607036517763\\
4	-0.063449278610827\\
5	-0.075512989071941\\
6	-0.0865236825214797\\
7	-0.0945730921906879\\
8	-0.10178273070777\\
9	-0.108070007533438\\
10	-0.113666759794782\\
11	-0.119042907205317\\
12	-0.124163054239821\\
13	-0.128724046438612\\
14	-0.133310043521288\\
15	-0.137564151504701\\
16	-0.142423826453301\\
17	-0.147380696324699\\
18	-0.152745250900429\\
19	-0.158640038607395\\
20	-0.164397781652687\\
21	-0.170483172181165\\
22	-0.17746916819977\\
23	-0.184947418083887\\
24	-0.193136808244703\\
25	-0.201861324164035\\
26	-0.211222590434302\\
27	-0.221217184117662\\
28	-0.23181145099262\\
29	-0.243006984676786\\
30	-0.254387982958646\\
31	-0.265860810158026\\
32	-0.276905654203445\\
33	-0.286958602712048\\
34	-0.295049196767406\\
35	-0.299538232353246\\
36	-0.297325975514717\\
37	-0.29096711593987\\
38	-0.280547382631716\\
39	-0.267121792680595\\
40	-0.251715817481992\\
41	-0.237761607613728\\
42	-0.225021805063027\\
43	-0.216690702841326\\
44	-0.215094876565651\\
};
\addlegendentry{front};

\addplot [color=red,solid]
  table[row sep=crcr]{%
1	4.2363633783019e-17\\
2	0.0124627449253655\\
3	0.0201325647828567\\
4	0.0245072137784886\\
5	0.0258488423021992\\
6	0.0257721546516129\\
7	0.0239677114270242\\
8	0.0214445429970804\\
9	0.0177212034327399\\
10	0.0134807113853668\\
11	0.00874497096590155\\
12	0.00366011607990218\\
13	-0.00189447647193277\\
14	-0.00766986587754141\\
15	-0.0139755101804077\\
16	-0.0206537799920005\\
17	-0.0278793474028525\\
18	-0.0356495618599542\\
19	-0.0437687595021773\\
20	-0.0525046967746347\\
21	-0.0619527555516213\\
22	-0.0722350926909885\\
23	-0.0832652556676756\\
24	-0.0954510265717551\\
25	-0.108926098760403\\
26	-0.123844332459264\\
27	-0.140410379562986\\
28	-0.159065133277505\\
29	-0.180183282585259\\
30	-0.204257789717002\\
31	-0.231862702366308\\
32	-0.263739677011365\\
33	-0.301083753198335\\
34	-0.345076646797909\\
35	-0.396965003265003\\
36	-0.451696592259976\\
37	-0.495205979508656\\
38	-0.52767806648344\\
39	-0.548849077773873\\
40	-0.557939833318892\\
41	-0.556437412222581\\
42	-0.541071785166004\\
43	-0.510600628102237\\
44	-0.469471853844793\\
};
\addlegendentry{rear};

\end{axis}
\end{tikzpicture}%
		\label{fig:steeringAngles}}
	\caption{Trajectory generation for a parking scenario}
	\label{fig_sim}
\end{figure}
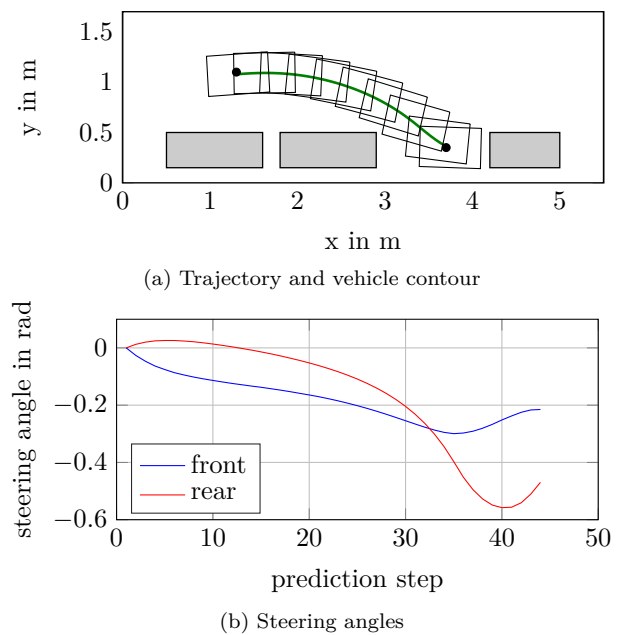
%
%
%
The following results have been obtained in simulation, using a model of the research vehicle MAX to show several scenarios for tactically chosen optimization parameters.
For an exemplary parking scenario, the results of trajectory generation and according steering angles are shown in Fig.~\ref{fig:parkingSzenario}.
The LPV model has proven to deal with slowly varying velocities during parking. 
The optimized trajectory is plotted in green, the vehicle contour is plotted  in black. 
It can be stated, that no collision occurs when following the calculated trajectory.
Also, vehicle heading is nearly zero for the final vehicle state in the parking lot. 
The trajectory is based on a steering configuration using counter steering and common steering, to achieve optimal results as shown in Fig.~\ref{fig:steeringAngles}. 

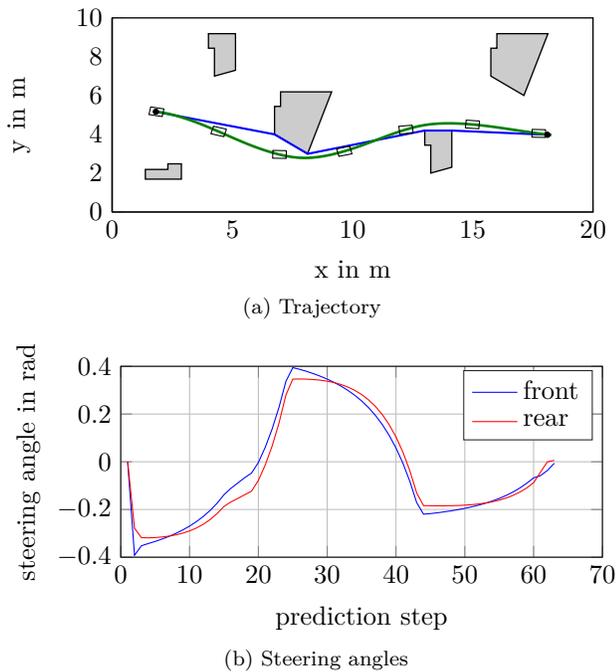
\begin{figure}[!h]
	\centering
	\subfloat[Trajectory]{
%
%
\definecolor{mycolor1}{rgb}{0.00000,0.44700,0.74100}%
\definecolor{mycolor2}{rgb}{1.00000,1.00000,0.00000}%
\definecolor{mycolor3}{rgb}{0.85000,0.32500,0.09800}%
\definecolor{mycolor4}{rgb}{0.40000,0.40000,0.90000}%
\begin{tikzpicture}

\begin{axis}[%
width=2.521in,
height=1.018in,
at={(0.677in,0.608in)},
scale only axis,
xmin=-0.02,
xmax=20.02,
xlabel={x in m},
ymin=-0.01,
ymax=10.01,
ylabel={y in m},
axis background/.style={fill=white},
axis x line*=bottom,
axis y line*=left
]

\addplot[area legend,solid,line width=0.5pt,draw=black,fill=white,forget plot]
table[row sep=crcr] {%
x	y\\
0	0\\
20	0\\
20	10\\
0	10\\
}--cycle;

\addplot[area legend,solid,line width=0.5pt,draw=black,fill=white!80!black,forget plot]
table[row sep=crcr] {%
x	y\\
1.375	1.6875\\
1.375	2.1875\\
2.3125	2.1875\\
2.3125	2.4767\\
2.8733	2.4767\\
2.8733	1.6875\\
}--cycle;

\addplot[area legend,solid,line width=0.5pt,draw=black,fill=white!80!black,forget plot]
table[row sep=crcr] {%
x	y\\
4.25	7\\
4.25	8.4375\\
4	8.4375\\
4	9.1875\\
5.125	9.1875\\
5.125	7.3\\
}--cycle;

\addplot[area legend,solid,line width=0.5pt,draw=black,fill=white!80!black,forget plot]
table[row sep=crcr] {%
x	y\\
6.75	4\\
6.75	5.4375\\
7	5.4375\\
7	6.1875\\
9.125	6.1875\\
8.125	3\\
}--cycle;

\addplot[area legend,solid,line width=0.5pt,draw=black,fill=white!80!black,forget plot]
table[row sep=crcr] {%
x	y\\
15.75	7\\
15.75	8.4375\\
16	8.4375\\
16	9.1875\\
18.125	9.1875\\
17.125	6\\
}--cycle;

\addplot[area legend,solid,line width=0.5pt,draw=black,fill=white!80!black,forget plot]
table[row sep=crcr] {%
x	y\\
13.25	2\\
13.25	3.4375\\
13	3.4375\\
13	4.1875\\
14.125	4.1875\\
14.125	2.3\\
}--cycle;
\addplot [color=black,mark size=1.0pt,only marks,mark=*,mark options={solid,fill=black,draw=black},forget plot]
  table[row sep=crcr]{%
1.8116	5.1636\\
};
\addplot [color=black,mark size=1.0pt,only marks,mark=*,mark options={solid,fill=black,draw=black},forget plot]
  table[row sep=crcr]{%
18.1022	3.9745\\
};
\addplot [color=blue,solid,line width=0.8pt,forget plot]
  table[row sep=crcr]{%
1.8116	5.1636\\
6.745	3.997\\
8.128	2.992\\
12.995	4.192\\
14.13	4.192\\
18.1022	3.9745\\
};
\addplot [color=black!50!green,solid,line width=1.0pt,forget plot]
  table[row sep=crcr]{%
1.81199893998956	5.16809133801171\\
2.08712587875953	5.11685076860652\\
2.35868497016141	5.05199226148799\\
2.62479075275909	4.96837769160392\\
2.88730860239072	4.87376673209048\\
3.14652728295407	4.76960858793401\\
3.40262446371776	4.65723733282743\\
3.65582533091363	4.53798931902281\\
3.90646391806269	4.41320525971246\\
4.15500257151643	4.28422916113465\\
4.40200773144874	4.15235870208312\\
4.64808703260478	4.01877424918541\\
4.89390154330098	3.88466141813778\\
5.14020344736737	3.75136636985278\\
5.38781872883716	3.62043148655896\\
5.63755496053953	3.49350178403283\\
5.89002046395988	3.37200206599907\\
6.14558977566332	3.25721189729409\\
6.4042867974498	3.14981064654376\\
6.66622359966137	3.05094667533414\\
6.93150903167203	2.96227984580501\\
7.20023832648315	2.88687167948027\\
7.47193221315007	2.82874132160466\\
7.74516505375396	2.79187828296638\\
8.01820840431922	2.77980868745586\\
8.29016005595919	2.79311246312216\\
8.56100070036024	2.82788467823636\\
8.8306438231292	2.87986584292009\\
9.09857217739069	2.94661904494879\\
9.36423692783241	3.0262033509614\\
9.62730731783627	3.11673173786082\\
9.88773605535875	3.21636517052594\\
10.1457492696885	3.3233169765247\\
10.4018077400349	3.43584573127289\\
10.6565169243343	3.55222578283744\\
10.9104851878259	3.67081476280808\\
11.1643658499218	3.79014391755008\\
11.418972614139	3.90853335553083\\
11.6751765284468	4.02407210101059\\
11.9338105138426	4.13471812369148\\
12.1956979586106	4.23827939871634\\
12.46172081382	4.33227714594781\\
12.7326823776293	4.41366987900903\\
13.0086665350449	4.47858978943802\\
13.2880746228796	4.52416151547819\\
13.5684055326464	4.55130400630235\\
13.8482778692006	4.56363796603776\\
14.127399462288	4.56402718724411\\
14.4057977420822	4.55413200529303\\
14.6835463891122	4.53534110952889\\
14.9607180080795	4.50886202587312\\
15.2373747964228	4.47578534039413\\
15.5135748612281	4.43713073578931\\
15.789382256593	4.3938789373174\\
16.0648728446462	4.34699477166861\\
16.3401494677191	4.29743413238762\\
16.6153291995324	4.24620074552145\\
16.8905430600011	4.19429143745096\\
17.1659359643896	4.14275447409222\\
17.4416562395115	4.09269523162022\\
17.7178445256187	4.04529117591322\\
17.9946178269744	4.00182948775313\\
18.2720226913657	3.96342959662663\\
};
\addplot [color=black,solid,forget plot]
  table[row sep=crcr]{%
2.148462509003	5.30886515207791\\
2.0752230411962	4.91562733554016\\
1.52469009804335	5.01816259046966\\
1.59792956585014	5.41140040700741\\
2.148462509003	5.30886515207791\\
};
\addplot [color=black,solid,forget plot]
  table[row sep=crcr]{%
4.74947243780591	4.26323375715019\\
4.64219621310836	3.87788740435229\\
4.1027113191913	4.02807411892886\\
4.20998754388885	4.41342047172676\\
4.74947243780591	4.26323375715019\\
};
\addplot [color=black,solid,forget plot]
  table[row sep=crcr]{%
7.24118374759483	3.15496240941191\\
7.23166144341305	2.75507576832088\\
6.6718201458856	2.76840699417538\\
6.68134245006739	3.16829363526641\\
7.24118374759483	3.15496240941191\\
};
\addplot [color=black,solid,forget plot]
  table[row sep=crcr]{%
9.87789878762414	3.38173912505891\\
9.97023920224648	2.99254348296181\\
9.42536530331054	2.86326690249053\\
9.3330248886882	3.25246254458763\\
9.87789878762414	3.38173912505891\\
};
\addplot [color=black,solid,forget plot]
  table[row sep=crcr]{%
12.4769916404652	4.47044068866085\\
12.5207441894211	4.07284074592964\\
11.9641042695974	4.01158717739132\\
11.9203517206415	4.40918712012253\\
12.4769916404652	4.47044068866085\\
};
\addplot [color=black,solid,forget plot]
  table[row sep=crcr]{%
15.2807231902176	4.68385254963022\\
15.2487625010788	4.28513145117946\\
14.6905529632478	4.32987641597368\\
14.7225136523865	4.72859751442443\\
15.2807231902176	4.68385254963022\\
};
\addplot [color=black,solid,forget plot]
  table[row sep=crcr]{%
18.0282038082395	4.23686913834903\\
18.017256759245	3.83701896376071\\
17.4574665148213	3.85234483235309\\
17.4684135638159	4.25219500694141\\
18.0282038082395	4.23686913834903\\
};
\addplot [color=black,solid,forget plot]
  table[row sep=crcr]{%
-0.656599951553432	-1.1228877810317\\
-0.975820497778523	-1.36392355196156\\
-1.31327057708033	-0.917014787246438\\
-0.994050030855244	-0.675979016316572\\
-0.656599951553432	-1.1228877810317\\
};
\addplot [color=black,solid,forget plot]
  table[row sep=crcr]{%
-0.660476597671561	-1.13322484479751\\
-0.986846101536034	-1.36448864859581\\
-1.31061542685365	-0.907571343185548\\
-0.984245922989179	-0.676307539387249\\
-0.660476597671561	-1.13322484479751\\
};
\addplot [color=black,solid,forget plot]
  table[row sep=crcr]{%
-0.662514719167208	-1.13830649016301\\
-0.99231899856214	-1.36464503591426\\
-1.30919296261389	-0.902919044761355\\
-0.979388683218954	-0.676580499010107\\
-0.662514719167208	-1.13830649016301\\
};
\addplot [color=black,solid,forget plot]
  table[row sep=crcr]{%
-0.663166327787158	-1.1398859437671\\
-0.994027285626763	-1.3646770169383\\
-1.30873478806644	-0.901471675962852\\
-0.977873830226837	-0.676680602791652\\
-0.663166327787158	-1.1398859437671\\
};
\end{axis}
\end{tikzpicture}%
		\label{fig:freeSzenario}}
	\hfil
	\vspace{-0.5em}
		\subfloat[Steering angles]{	
%
%
\definecolor{mycolor1}{rgb}{0.00000,0.44700,0.74100}%
\definecolor{mycolor2}{rgb}{0.85000,0.32500,0.09800}%
\begin{tikzpicture}

\begin{axis}[%
width=2.521in,
height=1.00in,
at={(0.813in,0.441in)},
scale only axis,
xmin=0,
xmax=70,
xlabel={prediction step},
xmajorgrids,
ymin=-0.4,
ymax=0.4,
ylabel={steering angle in rad},
ymajorgrids,
axis background/.style={fill=white},
legend style={legend cell align=left,align=left,draw=white!15!black}
]
\addplot [color=blue,solid]
  table[row sep=crcr]{%
1	1.08325753367861e-16\\
2	-0.391877686096854\\
3	-0.352052389199329\\
4	-0.342804257130443\\
5	-0.333873500933346\\
6	-0.323948326271176\\
7	-0.312814202627757\\
8	-0.300242576716133\\
9	-0.285955246713547\\
10	-0.269613331535888\\
11	-0.250801444793486\\
12	-0.229011481639844\\
13	-0.203629327745043\\
14	-0.173891737518855\\
15	-0.138192943898197\\
16	-0.111481394105394\\
17	-0.0893538397044186\\
18	-0.0669242163480204\\
19	-0.0475996891805246\\
20	-0.00303569512937957\\
21	0.0608629473296678\\
22	0.137457980144133\\
23	0.228711484566162\\
24	0.33680660978812\\
25	0.394361989536088\\
26	0.387383745036213\\
27	0.378782392409133\\
28	0.369179828024486\\
29	0.358439550610937\\
30	0.346362182388867\\
31	0.33269732892254\\
32	0.317137448254883\\
33	0.299308584895995\\
34	0.278758035665555\\
35	0.254935731171515\\
36	0.227168873614272\\
37	0.194642740865197\\
38	0.156382348474629\\
39	0.111217685407023\\
40	0.0577403480193876\\
41	-0.00574263767424776\\
42	-0.0812523668220942\\
43	-0.170609057355862\\
44	-0.219716066083736\\
45	-0.217305782291263\\
46	-0.213886089356741\\
47	-0.210007192815052\\
48	-0.205645190081208\\
49	-0.200739207598073\\
50	-0.195209901220022\\
51	-0.188957085930543\\
52	-0.181856113874837\\
53	-0.173753496903845\\
54	-0.164461581388101\\
55	-0.15375190358307\\
56	-0.141347934683677\\
57	-0.126915261581424\\
58	-0.110049532948834\\
59	-0.0902773662412594\\
60	-0.0675988284055832\\
61	-0.0580854522549536\\
62	-0.0365691668930691\\
63	-0.00623230842299291\\
};
\addlegendentry{front};

\addplot [color=red,solid]
  table[row sep=crcr]{%
1	-1.66845895705257e-17\\
2	-0.278506517153556\\
3	-0.318010232541841\\
4	-0.318899525714284\\
5	-0.317675713657515\\
6	-0.315360236826267\\
7	-0.311739952883431\\
8	-0.306527133015249\\
9	-0.299377663773106\\
10	-0.28988030877794\\
11	-0.277546292226979\\
12	-0.261793880791144\\
13	-0.241923566565621\\
14	-0.217118362591138\\
15	-0.187400274748602\\
16	-0.168540117886744\\
17	-0.153786962502545\\
18	-0.138435994587148\\
19	-0.123398573376943\\
20	-0.078383777444578\\
21	-0.0131624091801232\\
22	0.0660870732773153\\
23	0.162672442878366\\
24	0.279553041371742\\
25	0.346244434773889\\
26	0.346805558519048\\
27	0.346141144234641\\
28	0.344560341129663\\
29	0.34182059855176\\
30	0.337649708975738\\
31	0.331729503215636\\
32	0.323682201747335\\
33	0.313055699006795\\
34	0.299306990036946\\
35	0.281785950266939\\
36	0.259718711048258\\
37	0.232176235339539\\
38	0.198031623400474\\
39	0.155921324693354\\
40	0.104199853175206\\
41	0.0408792073089072\\
42	-0.0364453682508619\\
43	-0.130154137214353\\
44	-0.183707367254309\\
45	-0.184432705443658\\
46	-0.184592066012002\\
47	-0.184529331119572\\
48	-0.184140363093494\\
49	-0.183319899680839\\
50	-0.181946917335493\\
51	-0.17988008561373\\
52	-0.176952939545319\\
53	-0.172968070719848\\
54	-0.167690173910333\\
55	-0.160837891658039\\
56	-0.152073228541708\\
57	-0.14098986612636\\
58	-0.127099297827237\\
59	-0.109798659398569\\
60	-0.0877899057745863\\
61	-0.0434199929844828\\
62	-0.000382451714364626\\
63	0.00587333938069454\\
};
\addlegendentry{rear};

\end{axis}
\end{tikzpicture}%
		\label{fig:steeringAnglesFree}}
	\caption{Trajectory generation in an unstructured scenario}
	\label{fig_sim}
\end{figure}

Fig.~\ref{fig:freeSzenario} and Fig.~\ref{fig:steeringAnglesFree} show the results of a trajectory generation in an unstructured environment.
In Fig.~\ref{fig:freeSzenario}, the blue line segments depict the shortest path from start to target point, the collision-free trajectory is shown in green.  
Fig.~\ref{fig:steeringAnglesFree} shows the calculated steering angles to follow the generated trajectory. 
In this case, common steering is applied for the whole trajectory, to minimize yaw rate and thus increasing driving comfort. 
%
%
It can be stated, that trajectories for a variety of scenarios can be generated without changing optimization criteria. 
If high maneuverability is required, a trajectory based on counter steering is generated, otherwise common steering is applied to increase driving comfort.
This underlines the generic character of the presented approach.

\begin{figure}[h]
	\vspace{-0.2em}
	\centering
	\input{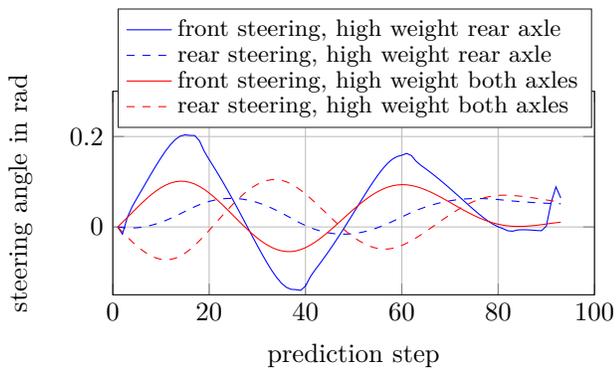}
	\vspace{-0.6em}
	\caption{Influencing trajectory by tuning weights}
	\label{fig:tuningWeights}
	\vspace{-0.6em}
\end{figure}

However, weight tuning can be applied if necessary.
Fig.~\ref{fig:tuningWeights} shows the effects of different weights applied for the steering angles at front and rear axle when planning a slalom maneuver. 
The blue lines result from optimization if a high weight on rear axle steering is applied. 
The tactical decision to choose a high weight for one axle could be performed  due to information from self-representation modules to reduce wear on the actuators of one axle. 
In case high weights are introduced for both axles the resulting steering angles switch to continuous counter steering, as this steering configuration requires lower steering angles to complete the slalom maneuver. 

The MPC framework runs with an average cycle time of ca. \SI{100}{\milli\second} on an Intel Pentium G3450 (\SI{3.4}{\giga\hertz}) using Mathworks' Embedded Coder for code generation.
\section{Conclusion \& Future Research}
\label{sec:conclusion}
In this paper, we presented an architectural approach to MPC for path planning in autonomous vehicles.
The approach considers the tactical formulation and reformulation of optimization parameters and constraints, based on self-assessment and feasibility information from the MPC framework.
The applicability of the chosen approach has been demonstrated in a simulative set-up for several planning scenarios.

Further research will focus on the integration of tire dynamics (e.g.\ as proposed by Yi et al. \cite{yi2016}).
Additionally we will extend the implementation of the self-aware MPC approach to vehicle dynamics control at the lower stabilization level (e.g.\ brake-blending).

Eventually, the presented control concepts shall be evaluated on the research platforms MOBILE and MAX.
\vspace{-0.8em}
\vspace{-0.5em}
\section*{Acknowledgment}
The authors would like to thank the MOBILE team, Leyre Tardio Martos in particular, for their valuable input.
This work is part of the DFG research group FOR1800 \emph{Controlling Concurrent Change} (CCC).
\vspace{-0.5em}
%
\bibliographystyle{IEEEtran}%
\bibliography{bib/references}%
\end{document}